\documentclass[amsmath,amssymb,prl]{revtex4}

\begin{document}
\title{Continuous variable noise-free states in correlated quantum noisy channels}
\author{Tohya Hiroshima}
\email{tohya@qci.jst.go.jp}
\affiliation{
Quantum Computation and Information Project, ERATO, Japan Science and Technology Agency,\\
Daini Hongo White Building 201, Hongo 5-28-3, Bunkyo-ku, Tokyo 113-0033, Japan
}
\author{Osamu Hirota}
\email{hirota@lab.tamagawa.ac.jp}
\affiliation{
Research Center for Quantum Communications,\\
Tamagawa University, Machida, 194-8610, Japan
}
%\date{\today}
%%%%%%%%%%%%%%%%%%%%%%%%%%%%%%%%%%%%%%%%%%%%%%%%%%%%%%%%%%%%%%%%%%%%%%%%%%%%%%%
\begin{abstract}
We explicitly compute the evolution of the density operator of a two-mode electromagnetic field when the influence of the thermal fluctuation of the vacuum is common for both modes.
From this result, we give an example in which the bundle of quantum noisy channels turns out to be noiseless for the special type of signal states due to the correlation.
\end{abstract}

\maketitle

Computing the classical capacity of quantum channels is one of the most important issues in quantum information theory \cite{NC}.
The quantum channels are represented by a stochastic or completely positive map that maps the density operator $\pi $ to the density operator $\sigma $; 
$\pi \mapsto \sigma =\Phi (\pi )$ \cite{Kraus}.
If the channels $\Phi _{1}$ and $\Phi _{2}$ have no correlation, the bundle of them is simply given by $\Phi _{1}\otimes \Phi _{2}$.
Even in this simple case, whether the classical capacity of quantum channels is additive or not is still a long-standing open question.
On the other hand, the correlation or memory effect between channels leads to interesting phenomena.
One such example was given in \cite{MP}, where the entangled signals enhance the mutual information of the pair of depolarizing channels with memory.
Without correlation, the depolarizing channels are proven to be additive \cite{King} and explicit calculations have shown that the input of entangled qubits cannot increase the classical capacity of depolarizing channels \cite{BFMP}.
Similar results for channels with memory have been obtained in \cite{MPV,BDB,YS}.
Asymptotic classical capacity of quantum channels with memory has been also analyzed \cite{BM}.

In this paper, we give an example in which the bundle of quantum noisy channels turns out to be noiseless for the special type of signal states due to the correlation.
This constitutes one example of the noiseless codes \cite{PSE,KBLW} in a continuous variable system.

Consider a two-mode electromagnetic field with the same frequency $\omega _{0}$ interacting with the reservoir: the thermal fluctuation of the vacuum.
If these two modes are spatially well separated, the reservoir field acts independently on each mode.
The interaction Hamiltonian takes the following form.
\begin{equation} \label{eq:Interaction1}
V=\hbar \sum_{i=1}^{2}a_{i}^{\dagger }\sum_{j}g_{j}^{(i)}b_{j}^{(i)}+h.c..
\end{equation}
Here, $a_{i}^{\dagger }$ denotes the creation operator of the electromagnetic field of mode $i$ and $b_{j}^{(i)}$ denotes the annihilation operator of the electromagnetic field of reservoir $i$ acting on mode $i$.
They obey the commutation relation $\left[ a_{i},a_{j}^{\dagger }\right] _{-}=\delta _{ij}$ and $\left[ b_{k}^{(i)},b_{l}^{(j)\dagger }\right] _{-}=\delta _{ij}\delta _{kl}$.
Other commutators are zero.
$g_{j}^{(i)}$ is the coupling constant.
This independent reservoir model has been well investigated \cite{JLK,Hiroshima,SW01,WJK}.
However, if the two modes are close enough spatially, the action of noise on each mode must be more or less correlated.
More precisely, the $i$th reservoir field $b^{(i)}$ acts partially on mode $j(\neq i)$.
Therefore, $b_{j}^{(i)}$ in Eq.~(\ref{eq:Interaction1}) should be replaced by 
$\sqrt{1-\mu }b_{j}^{(i)}+\sqrt{\mu }b_{j}$ with 
$\left[ b_{i},b_{j}^{\dagger }\right] _{-}=\delta _{ij}$.
Here, $b_{j}$ is the annihilation operator of the common reservoir that is independent of reservoirs 1 and 2 and the parameter $\mu$ ($0\leq \mu \leq 1$) represents the degree of correlation.
Under the extreme condition of $\mu \rightarrow 1$, the interaction Hamiltonian takes the form
\begin{equation} \label{eq:Interaction2}
V=\hbar \left( \sum_{i=1}^{2}a_{i}^{\dagger }\right) \sum_{j}g_{j}b_{i}+h.c.,
\end{equation}
where $g_{j}=g_{j}^{(1)}=g_{j}^{(2)}$.
Namely, the reservoir acts on both modes simultaneously.
The same model has been applied to the two-mode squeezed vacuum states \cite{P-B}.
This ideal model may be hard to achieve experimentally, but it provides the essential feature of these correlated channels of intermediate value of $\mu$ ($0 < \mu < 1$).
Hereafter, we confine ourselves to this extreme case.

Following the standard procedure for deriving the master equation for the density operator under the Markov approximation in the weak coupling limit \cite{WM}, we get the following equation of motion for the two-mode density operator in the interaction picture.
\begin{eqnarray} \label{eq:MasterEquation}
\frac{d}{dt}\rho (t) &=&\frac{\Gamma }{2}(N_{0}+1)\left[ 2(a_{1}+a_{2})\rho
(t)(a_{1}^{\dagger }+a_{2}^{\dagger })-(a_{1}^{\dagger }+a_{2}^{\dagger
})(a_{1}+a_{2})\rho (t)-\rho (t)(a_{1}^{\dagger }+a_{2}^{\dagger
})(a_{1}+a_{2})\right]   \nonumber \\
&&+\frac{\Gamma }{2}N_{0}\left[ 2(a_{1}^{\dagger }+a_{2}^{\dagger })\rho
(t)(a_{1}+a_{2})-(a_{1}+a_{2})(a_{1}^{\dagger }+a_{2}^{\dagger })\rho
(t)-\rho (t)(a_{1}+a_{2})(a_{1}^{\dagger }+a_{2}^{\dagger })\right] ,
\end{eqnarray}
where 
$N_{0}=1/\left( e^{\hbar \omega _{0}/kT}-1\right) $ 
is the mean number of reservoir quanta at frequency $\omega _{0} $.
The initial ($t=0$) density operator can be expanded in terms of coherent states as follows.
\begin{equation} \label{eq:InitialRho}
\rho (0) =\int \cdots \int \prod_{i=1}^{2}d^{2}\alpha _{i}d^{2}\beta _{i}
P\left( \alpha _{1},\alpha _{2};\beta _{1}^{*},\beta _{2}^{*}\right)
\Lambda \left( \alpha _{1},\alpha _{2};\beta _{1}^{*},\beta _{2}^{*}\right),
\end{equation}
where
\begin{equation}
\int \cdots \int \prod_{i=1}^{2}d^{2}\alpha _{i}d^{2}\beta _{i}P\left(
\alpha _{1},\alpha _{2};\beta _{1}^{*},\beta _{2}^{*}\right) =1,
\end{equation}
and
\begin{equation}
\Lambda \left( \alpha _{1},\alpha _{2};\beta _{1}^{*},\beta _{2}^{*}\right) =
\frac{\left| \alpha _{1},\alpha _{2}\right\rangle
\left\langle \beta _{1},\beta _{2}\right| }{\left\langle \beta _{1}\right|
\left. \alpha _{1}\right\rangle \left\langle \beta _{2}\right| \left. \alpha _{2}\right\rangle }
=\prod\limits_{i=1}^{2}\exp \left( -\alpha _{i}\beta _{i}^{*}\right)
\exp(\alpha _{i}a_{i}^{\dagger })\left| 0\right\rangle \left\langle 0\right|
\exp (\beta _{i}^{*}a_{i})
\end{equation}
is an analytic function of $\alpha _{1}$, $\alpha _{2}$, $\beta _{1}^{*}$, and $\beta_{2}^{*}$.
$\rho (t)$ is the output state for the input state $\rho (0)$ in the channel with correlation considered here, $\Phi _{c}$; $\rho (t)=\Phi _{c}\left[ \rho (0)\right] $.

In order to solve Eq.~(\ref{eq:MasterEquation}) under the initial condition (\ref{eq:InitialRho}), we first calculate the two-mode characteristic function defined as
\begin{equation} \label{eq:Characteristic}
\chi (\lambda _{1},\lambda _{2};t)={\rm Tr}\left[ \prod_{i=1}^{2}\exp
\left( \lambda _{i}a_{i}^{\dagger }-\lambda _{i}^{+}a_{i}\right) \rho
(t)\right]. 
\end{equation}
Following the standard procedure \cite{WM}, we get the equation of motion for $\chi (\lambda _{1},\lambda _{2};t)$ from Eqs.~(\ref{eq:MasterEquation}) and (\ref{eq:Characteristic}) as
\begin{equation} \label{eq:CharacteristicEquation}
\frac{d}{dt}\chi (\lambda _{1},\lambda _{2};t)=
-\Gamma \left( N_{0}+\frac{1}{2}\right) \left| \lambda _{+}\right|^{2}
\chi (\lambda _{1},\lambda _{2};t)
-\Gamma \left( \lambda _{+}^{*}\frac{\partial }{\partial \lambda _{+}^{*}}
+\lambda _{+}\frac{\partial }{\partial \lambda _{+}}\right)
\chi (\lambda _{1},\lambda _{2};t),
\end{equation}
where 
$\lambda _{\pm }=\lambda _{1}\pm \lambda _{2}$.
From Eqs.~(\ref{eq:InitialRho}) and (\ref{eq:Characteristic}), the initial value of $\chi (\lambda _{1},\lambda _{2};t)$ is calculated as
\begin{equation} \label{eq:CharacteristicInitial}
\chi (\lambda _{1},\lambda _{2};0) =\int \cdots \int
\prod_{i=1}^{2}d^{2}\alpha _{i}d^{2}\beta _{i}
P\left( \alpha _{1},\alpha _{2};\beta _{1}^{*},\beta _{2}^{*}\right)
\prod\limits_{i=1}^{2}\exp 
\left( -\frac{\left| \lambda _{i}\right|^{2}}{2}-\lambda _{i}^{*}\alpha _{i}+\lambda _{i}\beta _{i}^{*}\right).
\end{equation}
Following the operator method described in \cite{Collett}, we can solve Eq.~(\ref{eq:CharacteristicEquation}) under the initial condition (\ref{eq:CharacteristicInitial}).
The straightforward calculations yield
\begin{eqnarray}
&&\chi (\lambda _{1},\lambda _{2};t)=\int \cdots \int
\prod_{i=1}^{2}d^{2}\alpha _{i}d^{2}\beta _{i}P\left( \alpha _{1},\alpha
_{2};\beta _{1}^{*},\beta _{2}^{*}\right)   \nonumber \\
&&\times \exp \left[ -\frac{1}{2}N(t)\left| \lambda _{+}\right| ^{2}-\frac{1%
}{4}\left| \lambda _{+}\right| ^{2}-\frac{1}{4}\left| \lambda _{-}\right|
^{2}-\frac{1}{2}e^{-\Gamma t}\left( \alpha _{+}\lambda _{+}^{*}-\beta
_{+}^{*}\lambda _{+}\right) -\frac{1}{2}\left( \alpha _{-}\lambda
_{-}^{*}-\beta _{-}^{*}\lambda _{-}\right) \right] ,
\end{eqnarray}
with 
$\alpha _{\pm }=\alpha _{1}\pm \alpha _{2}$, 
$\beta _{\pm }=\beta _{1}\pm \beta _{2}$, 
and 
$N(t)=N_{0}(1-e^{-2\Gamma t})$.
It is an easy task to check that $\chi (\lambda _{1},\lambda _{2};t)$ obtained above satisfies Eqs.~(\ref{eq:CharacteristicEquation}) and (\ref{eq:CharacteristicInitial}).

In the following, we calculate the two-mode Q-function from 
$\chi (\lambda _{1},\lambda _{2};t)$ 
and then calculate the density operator from the two-mode Q-function, which is defined as the diagonal element of the density operator in a coherent state; 
$ Q(\delta _{1},\delta _{2};t)=
\left\langle \delta _{1},\delta _{2}\right| \rho (t)\left| \delta _{1},\delta _{2}\right\rangle 
/ \pi ^{2}$.
This is also calculated from the two-mode anti-normally ordered characteristic function
\begin{equation}
\chi ^{A}(\lambda _{1},\lambda _{2};t)={\rm Tr}\left[ \prod_{i=1}^{2}\exp
(-\lambda _{i}^{*}a_{i})\exp (\lambda _{i}a_{i}^{\dagger })\rho (t)\right]
\end{equation}
via the following integral transformation
\begin{equation}
Q(\delta _{1},\delta _{2};t) =
\frac{1}{\pi ^{4}}\int \int \prod_{i=1}^{2}d^{2}\lambda _{i}
\exp \left( \delta _{i}\lambda _{i}^{*}-\delta _{i}^{*}\lambda _{i}\right)
\chi ^{A}(\lambda _{1},\lambda _{2};t).
\end{equation}
Since 
$\chi ^{A}(\lambda _{1},\lambda _{2};t)=e^{-\left| \lambda _{1}\right|
^{2}/2-\left| \lambda _{2}\right| ^{2}/2}\chi (\lambda _{1},\lambda _{2};t)$, 
we can calculate the Q-function by using the identity \cite{WM,Puri}
\begin{equation} \label{eq:Identity}
\frac{1}{\pi }\int d^{2}\eta \exp \left( -\lambda \left| \eta \right|
^{2}+\mu \eta +\nu \eta ^{*}\right) =\frac{1}{\lambda }\exp \left( \frac{\mu
\nu }{\lambda }\right),
\end{equation}
which holds for 
$\mathrm{{Re}(\lambda ) > 0}$ 
and arbitrary $\mu$ and $\nu$.
The result is
\begin{equation} \label{eq:Qfunction}
Q(\delta _{1},\delta _{2};t)=\frac{1}{\pi ^{2}}\frac{1}{1+N(t)}\int \cdots
\int \prod_{i=1}^{2}d^{2}\alpha _{i}d^{2}\beta _{i}P\left( \alpha
_{1},\alpha _{2};\beta _{1}^{*},\beta _{2}^{*}\right) \exp \left[ \frac{1}{2}%
f(\alpha _{1},\alpha _{2};\beta _{1}^{*},\beta _{2}^{*};\delta _{1},\delta
_{2})\right] ,
\end{equation}
where
\begin{eqnarray} \label{eq:QfunctionExponential}
&&f(\alpha _{1},\alpha _{2};\beta _{1}^{*},\beta _{2}^{*};\delta _{1},\delta _{2})  \nonumber \\
&=&-\frac{2+N(t)}{1+N(t)}
\left( \left| \delta _{1}\right|^{2}+\left| \delta _{2}\right| ^{2}\right)
+\frac{N(t)}{1+N(t)}\left( \delta _{1}\delta _{2}^{*}+\delta _{1}^{*}\delta _{2}\right)
-\frac{1}{1+N(t)}\alpha _{+}\beta _{+}^{*}e^{-2\Gamma t}
-\alpha _{-}\beta _{-}^{*} \nonumber \\
&&+\left( \frac{\alpha _{+}e^{-\Gamma t}}{1+N(t)}+\alpha _{-}\right) \delta _{1}^{*}
+\left( \frac{\beta _{+}^{*}e^{-\Gamma t}}{1+N(t)}+\beta _{-}^{*}\right) \delta _{1}
+\left( \frac{\alpha _{+}e^{-\Gamma t}}{1+N(t)}-\alpha _{-}\right) \delta _{2}^{*}
+\left( \frac{\beta _{+}^{*}e^{-\Gamma t}}{1+N(t)}-\beta _{-}^{*}\right) \delta _{2}.
\end{eqnarray}
Now, we get the density operator from the following formula.
\begin{equation} \label{eq:DensityMatrix}
\rho (t) =\left( \frac{1}{4\pi }\right) ^{2}\int \cdots \int
\prod_{i=1}^{2}d^{2}\delta _{i}d^{2}\gamma _{i}\exp \left( -\frac{1}{4}%
\left| \delta _{i}-\gamma _{i}\right| ^{2}\right)
\Lambda \left( \delta _{1},\delta _{2};\gamma _{1}^{*},\gamma
_{2}^{*}\right)
Q\left( \frac{1}{2}\left( \delta _{1}+\gamma _{1}\right) ,\frac{1}{2%
}\left( \delta _{2}+\gamma _{2}\right) ;t\right).
\end{equation}
This is an obvious extension of the relation \cite{DG}
\begin{equation} \label{eq:Formula}
\rho =\frac{1}{\pi }\int \int d^{2}\gamma d^{2}\delta \Lambda (\gamma
+\delta ,\gamma -\delta )e^{-\left\vert \delta \right\vert ^{2}}Q(\gamma )
\end{equation}
to the two-mode case.
In Eq.~(\ref{eq:Formula}), 
$\Lambda (\alpha ,\beta )=\left| \alpha \right\rangle \left\langle \beta
^{*}\right| /\left\langle \beta ^{*}\right| \left. \alpha \right\rangle $ 
and 
$Q(\gamma )=\left\langle \gamma \right| \rho \left| \gamma \right\rangle /\pi $ 
is the single-mode Q-function.

At zero temperature, i.e., $N_{0}=0$, the integral of Eq.~(\ref{eq:DensityMatrix}), can be explicitly performed by using the identity
\begin{equation}
\frac{1}{\pi }\int d^{2}\alpha \exp \left( -\left| \alpha \right| ^{2}+\beta
\alpha ^{*}\right) f(\alpha )=f(\beta )
\end{equation}
for an arbitrary analytic function $f$.
The result is
\begin{equation}
\rho (t) =\int \cdots \int \prod_{i=1}^{2}d^{2}\alpha _{i}d^{2}\beta
_{i}P\left( \alpha _{1},\alpha _{2};\beta _{1}^{*},\beta _{2}^{*}\right) 
\Lambda \left( \alpha _{+}(t),\alpha _{-}(t);\beta
_{+}^{*}(t),\beta _{-}^{*}(t)\right),
\end{equation}
where 
$\alpha _{\pm }(t)=(\alpha _{+}e^{-\Gamma t}\pm \alpha _{-})/2$ 
and 
$\beta _{\pm }(t)=(\beta _{+}e^{-\Gamma t}\pm \beta _{-})/2$.
It is quite easy to see that $\rho (t)$ thus obtained satisfies the master equation (\ref{eq:MasterEquation}) with $N_{0}=0$.
It should be noted that the initial pure state 
$\rho (0)=\left| \alpha _{1},\alpha _{2}\right\rangle \left\langle \alpha
_{1},\alpha _{2}\right| $ 
remains pure; it becomes 
$\rho (t)=\Phi _{c}\left[ \rho (0)\right]=\left| \alpha _{+}(t),\alpha _{-}(t)\right\rangle \left\langle \alpha _{+}(t),\alpha _{-}(t)\right| $ 
as the output of the channel $\Phi _{c}$.
More interestingly, the state 
$\left| \alpha ,-\alpha \right\rangle \left\langle \alpha ,-\alpha
\right| $ is not affected by the noise of the channel $\Phi _{c}$.
This is also true for the entangled coherent state \cite{Sanders,vH} of the form $\left( \left| \alpha ,-\alpha \right\rangle \pm e^{i\phi }\left| -\alpha ,\alpha \right\rangle \right) /\sqrt{2}$.
These perfectly decoherence-free states arise from the quantum interference coming from the indistinguishability of modes 1 and 2 for the common reservoir.
For the input state $\rho (0)=\left| \alpha ,-\alpha \right\rangle \left\langle \alpha ,-\alpha
\right| $, the quantum interference is completely constructive while for $\rho (0)=\left| \alpha ,\alpha \right\rangle \left\langle \alpha ,\alpha \right| $, it is destructive.

At finite temperatures $T>0$, the input pure state does not keep its purity during the time evolution: the output state is mixed even if the input state is pure.
The output purity ${\rm Tr}\rho ^{2}(t)$ is calculated from Eq.~(\ref{eq:DensityMatrix}) with Eqs.~(\ref{eq:Qfunction}) and (\ref{eq:QfunctionExponential}) by using identity (\ref{eq:Identity}).
The explicit calculations yield
\begin{equation}
{\rm Tr}\rho ^{2}(t)=\frac{2\left( 1+N(t)\right) }{2+6N(t)+5N(t)^{2}}.
\end{equation}
This is a monotonically decreasing function of $t$.
Another measure for characterizing the channel is the fidelity that quantifies how close the output state $\rho (t)$ and the input state $\rho (0)=\left\vert \alpha _{1},\alpha _{2}\right\rangle \left\langle \alpha _{1},\alpha _{2}\right\vert $ are:
\begin{equation}
F\left( \alpha _{1},\alpha _{2}\right) ={\rm Tr}\left[ \rho (0)\rho (t)%
\right]. 
\end{equation}
This is also calculated from Eq.~(\ref{eq:DensityMatrix}) with Eqs.~(\ref{eq:Qfunction}) and (\ref{eq:QfunctionExponential}) by using identity (\ref{eq:Identity}).
The fidelity for the initial pure state $\rho (0)=\left| \alpha ,-\alpha \right\rangle \left\langle \alpha ,-\alpha \right| $ is
\begin{equation}
F\left( \alpha,-\alpha\right) =\frac{1}{1+N(t)},
\end{equation}
that is independent of $\alpha$ and tends to 1 for $N_{0}\rightarrow 0$.
This clearly contrasts with the fidelity
\begin{equation}
F\left( \alpha,\alpha\right) =\frac{1}{1+N(t)}\exp 
\left[ -\frac{2(1-e^{-\Gamma t})^{2}}{1+N(t)}\left| \alpha \right| ^{2}\right]
\end{equation}
calculated for the initial state 
$\rho (0)=\left| \alpha ,\alpha \right\rangle \left\langle \alpha ,\alpha \right| $.
The state $\left| \alpha ,-\alpha \right\rangle \left\langle \alpha ,-\alpha \right| $ is almost noise-free even for finite temperatures while the state $\left| \alpha ,\alpha \right\rangle \left\langle \alpha ,\alpha \right| $ is much more influenced by the noise.

In summary, we gave an example of the noise-free or decoherence-free states in quantum noisy channels with correlation by analyzing two electromagnetic fields coupled to a common thermal reservoir.
The results suggest that we can construct a partially noise-free signal state in the bundle of electromagnetic field channels if the modes of signal field are spatially close enough to share partially the common thermal reservoir.


\begin{thebibliography}{99}

\bibitem{NC}  M. A. Nielsen and I. L. Chuang, {\it Quantum Computation and Quantum Information} (Cambridge University Press, Cambridge, 2000).

\bibitem{Kraus}  K. Kraus, {\it States, Effects, and Operations: Fundamental Notions of Quantum Theory} (Springer-Verlag, Berlin, 1983).

\bibitem{MP}  C. Macchiavello and G. M. Palma, Phys. Rev. A {\bf 65}, 050301(R) (2002).

\bibitem{King}  C. King, IEEE Trans. Inf. Theory {\bf 49}, 221 (2003).

\bibitem{BFMP}  D. Bruss, L. Faoro, C. Macchiavello, and G. M. Palma, J. Mod. Opt. {\bf 47}, 325 (2000).

\bibitem{MPV}  C. Macchiavello, G. M. Palma, and S. Virmani, Phys. Rev. A {\bf 69}, 010303(R) (2004).

\bibitem{BDB}  J. Ball, A. Dragan, and K. Banaszek, quant/ph-0309148.

\bibitem{YS}  Y. Yeo and A. Skeen. Phys. Rev. A {\bf 67}, 064301 (2003).

\bibitem{BM}  G. Bowen and S. Mancini, Phys. Rev. A {\bf 69}, 012306 (2004).

\bibitem{PSE}  G. M. Palma, K.-A. Suominen, and A. K. Ekert, Proc. R. Soc. London, Ser. A {\bf 452}, 567 (1996).

\bibitem{KBLW}  J. Kempe, D. Bacon, D. A. Lidar, and K. B. Whaley, Phys. Rev. A {\bf 63}, 042307 (2001).

\bibitem{JLK}  H. Jeong, J. Lee, and M. S. Kim, Phys. Rev. A {\bf 61}, 052101 (2000).

\bibitem{Hiroshima}  T. Hiroshima, Phys. Rev. A {\bf 63}, 022305 (2001).

\bibitem{SW01}  S. Scheel and D.-G. Welsch, Phys. Rev. A {\bf 64}, 063811 (2001).

\bibitem{WJK}  D. Wilson, H. Jeong, and M. S. Kim, J. Mod. Opt. {\bf 49}, 851 (2002).

\bibitem{P-B}  J. S. Prauzner-Bechcicki, J. Phys. A {\bf 37}, L173 (2004).

\bibitem{WM}  D. F. Walls and G. J. Milburn, {\it Quantum Optics} (Springer-Verlag, Berlin, 1994).

\bibitem{Collett}  M. J. Collett, Phys. Rev. A {\bf 38}, 2233 (1988).

\bibitem{Puri}  R. R. Puri, {\it Mathematical Methods of Quantum Optics} (Springer-Verlag, Berlin, 2001).

\bibitem{DG}  P. D. Drummond and C. W. Gardiner, J. Phys. A {\bf 13}, 2353 (1980).

\bibitem{Sanders}  B. C. Sanders, Phys. Rev. A {\bf 45}, 6811 (1992).

\bibitem{vH}  S. J. van Enk and O. Hirota, Phys. Rev. A {\bf 64}, 022313 (2001).

\end{thebibliography}
\end{document}